">STUDENT AGENCY IN AI-ASSISTED LEARNING (PREPRINT)# A Theoretical Framework of Student Agency in AI-Assisted Learning: A Grounded Theory Approach

">**Yun Dai, Sichen Lai**

The Chinese University of Hong Kong## Abstract

abstract">Generative AI (GenAI) is a kind of AI model capable of producing human-like content in various modalities, including text, image, audio, video, and computer programming. Although GenAI offers great potential for education, its value often depends on students' ability to engage with it actively, responsibly, and critically—qualities central to student agency. Nevertheless, student agency has long been a complex and ambiguous concept in educational discourses, with few empirical studies clarifying its distinct nature and process in AI-assisted learning environments. To address this gap, the qualitative study presented in this article examines how higher education students exercise agency in AI-assisted learning and proposes a theoretical framework using a grounded theory approach. Guided by agentic engagement theory, this article analyzes the authentic experiences of 26 students using data from their GenAI conversation records and cognitive interviews that capture their thought processes and decision-making. The findings identify four key aspects of student agency: *initiating and (re)directing*, *mindful adoption*, *external help-seeking*, and *reflective learning*. Together, these aspects form an empirically developed framework that characterizes student agency in AI-assisted learning as a proactive, intentional, adaptive, reflective, and iterative process. Based on the empirical findings, theoretical and practical implications are discussed for researchers, educators, and policymakers.

**Keywords:** artificial intelligence, generative AI, agency, agentic engagement, AI-assisted learning, grounded theory">**A Preprint Version**:

Dai, Y., & Lai, S. (2025). A Theoretical Framework of Student Agency in AI-Assisted Learning: A Grounded Theory Approach. In L. Khan (Ed.), *Oxford Intersections: Social Media in Society and Culture* (pp. 0). Oxford University Press.

# Introduction

Generative AI (GenAI) is a kind of AI model capable of producing human-like content in various modalities, including text, image, audio, video, and computer programming. Since the launch of tools such as ChatGPT, higher education students have increasingly adopted GenAI tools to assist their learning tasks, including writing, coding, data analysis, and receiving feedback (Lai et al., 2025; Rahman & Watanobe, 2023). This rapid uptake, however, has raised concerns about GenAI's inherent limitations (e.g., inconsistent content quality and a lack of creativity and originality) and ethical issues such as misuse and academic integrity violations (Stahl & Eke, 2024). These challenges call for students to move beyond passive use and adopt a critical, reflective, and agentic stance when engaging with AI-generated content (Dai et al., 2023a).

This emphasis on active engagement with GenAI brings student agency to the forefront of educational inquiry (Darvishi et al., 2024). Student agency refers to the capacity to take ownership of learning, make autonomous decisions about the use of resources such as GenAI tools, and actively shape one's learning experiences (Archer, 2000). However, empirical understanding of how students exercise agency in AI-assisted learning remains limited (Yang et al., 2024). This gap is partly due to the conceptual complexity of agency, which challenges efforts to define, operationalize, and observe it in real-world educational settings (Matusov et al., 2016). Existing research often relies on classical theories or proxy constructs such as self-determination and self- efficacy rather than directly examining student agency as it emerges in practice. There is a need for grounded, context-sensitive approaches that capture how agency is enacted and manifested within contemporary AI- assisted learning environments.

Against this background, this article proposes an empirically developed framework for student agency in AI- assisted learning through a grounded theory approach. Informed by literature on AI-assisted learning and agentic engagement, we systematically examine how students exercise agency in real-world interactions with GenAI tools for academic purposes. The article draws on qualitative data from cognitive interviews and records of student–AI interactions, which allows us to capture the nuanced processes and strategies students employ. Our aim is to identify and conceptualize the core dimensions of student agency as they emerge in AI-assisted learning. The findings contribute to the theoretical understanding of learner agency in the context of emerging technologies and offer practical insights for educators seeking to support agency in AI-integrated learning environments.

## Literature Review

### AI-Assisted Learning

AI-assisted learning, an emerging area within learning sciences and technologies, refers to the integration of AI technologies into the educational process to enhance the efficiency, effectiveness, and personalization of learning experiences (Vargas-Murillo et al., 2023; Wang et al., 2023). Theoretically, it represents a convergence of cognitive science, computer science, and pedagogical theory, wherein AI technologies are employed as a source of assistance to support and extend human learning capabilities (Pham et al., 2023). These technologies can provide adaptive feedback, automate routine tasks, facilitate complex problem-solving, and offer tailored learning pathways based on individual needs and performance patterns (Dai et al., 2023a). It can be found in a range of educational contexts, including kindergarten through grade 12 and higher education, as well as professional and adult training (Dai et al., 2023b; Diliberti et al., 2024; Langran et al., 2024; Milana et al., 2024). AI-assisted learning is not confined to embedded learning experiences within specifically developed intelligent platforms such as intelligent tutoring systems and adaptive learning environments. It also encompasses students' voluntary adoption of external AI applications, such as ChatGPT and Claude, in their learning activities (Le et al., 2013).

The use of AI technologies to assist student learning dates back to the 1970s, and the field of AI-assisted learning has evolved in parallel with advances in AI research and computing power. In the 1970s, the most representative application was early intelligent tutoring systems, which used rule-based logic to interpret student responses and provide step-by-step guidance (Nwana, 1990). Although basic, they offered a first step toward personalization. Since the 2000s, advances in machine learning, data mining, and predictive analytics have enabled systems to analyze large data sets of learner interactions (Lang et al., 2017; Russell & Norvig, 2021). This led to the development of learning analytics tools that could identify misconceptions, predict outcomes, and flag dropout risks, allowing for earlier and more targeted interventions (Clow, 2013). Since the early 2020s, the emergence of GenAI, particularly large language models, has marked a new phase. These systems can produce human-like language, create varied content on demand, and engage in open-ended dialogue, greatly expanding AI-assisted learning beyond specialized platforms to widely accessible tools such as ChatGPT and Claude (Holmes & Miao, 2023).

Compared with earlier intelligent systems, GenAI offers distinct capabilities for learning. It can produce a wide range of outputs—such as essays, code, and statistical analyses—based on student prompts, providing rapid, on-demand content generation. Although output quality may vary, this functionality offers significant convenience and efficiency for students in their learning activities (Rudolph et al., 2023). GenAI also enables interactive, conversational engagement. Many tools use prompt-based interfaces, where students give instructions—called prompts—in natural language or another input format (Mayer et al., 2023). These exchanges are responsive and adaptive, as the AI adjusts its output dynamically to student inputs, available data, and emerging patterns. This adaptability supports personalized, iterative interactions between the learner and the system, allowing students to explore topics in depth, resolve emerging confusions, and receive tailored support in real time (Baidoo-Anu & Owusu Ansah, 2023).

**Student Agency in AI-Assisted Learning**

The affordances of GenAI tools have direct implications for student agency: They can expand opportunities for choice, control, and self-directed learning, but they may also risk fostering dependence if students offload too much cognitive effort to the system (Darvishi et al., 2024; Yang et al., 2024). Dai et al. (2023a) conceptualized student engagement with AI as a student-centered and student-driven process, in which learners actively shape the interaction through their prompts. By deciding what to ask, how to frame their requests, and how to respond to AI outputs, students influence both the direction and the quality of the learning exchange. This means that GenAI can support rich, meaningful dialogue and deeper learning when students engage thoughtfully, but it may result in shallow, less productive interactions when engagement is minimal. In this way, the degree of student agency becomes a decisive factor in determining the effectiveness of AI-assisted learning.

In addition, the inherent limitations and ethical implications of GenAI tools highlight the irreplaceable role of learner agency in AI-assisted learning. GenAI outputs can be inaccurate, biased, or culturally narrow, often lacking transparency in how responses are generated (Bontcheva et al., 2024; Rudolph et al., 2023). Such limitations require learners to critically evaluate and verify information rather than accept it uncritically (Dai et al., 2023a). The ease of producing high-quality text or solutions also raises challenges to academic integrity and authorship, as overreliance on AI may obscure the learner's original contribution and weaken their sense of ownership over the work (Michel-Villarreal et al., 2023 ). Without deliberate engagement, students risk bypassing the cognitive effort necessary for deep learning, reducing opportunities to develop higher order thinking skills (Yang et al., 2024). Cultivating mindful, reflective use of GenAI—where learners question, adapt, and integrate AI outputs—is therefore essential. From this perspective, student agency serves as a safeguard against the pitfalls of GenAI and a foundation for cultivating responsible, thoughtful learners who can thrive in an increasingly automated world (Chan & Hu, 2023; Dai, 2025).

Despite its significance, student agency in AI-assisted learning remains an ambiguous and underdefined construct, reflecting a long-standing lack of conceptual clarity in broader literature (Archer, 2000; Matusov et al., 2016). A scoping review by Roe and Perkins (2024) identifies three major themes in current literature: control in digital spaces, variable engagement and access, and changing notions of agency. While these themes reflect the diversity of educational discourses, early studies document shifting ways in which learners interact with technologies and exercise critical capabilities in engaging with GenAI tools (Yan et al., 2024). However, most existing studies often rely on alternative indicators—such as comment length, content relatedness, or the number of likes in peer feedback (Darvishi et al., 2024)—or broad proxies such as learner satisfaction and self- regulation (Marín et al., 2020). Such measures fail to capture the unique, situated, and evolving dynamics of human–AI interaction. As such, there is a need for a grounded, bottom-up approach that directly observes agency as it manifests in context, laying a foundation for developing an empirically grounded conceptual framework.

**Theoretical Framework: Agentic Engagement**

Human agency has long been regarded as a tricky concept due to the difficulty in defining, operationalizing, and measuring it (Archer, 2000). One reason is that scholars have not yet reached a consensus about agency, especially in educational contexts (Hitlin & Elder, 2007). Since the mid-2000s, there has been a growing emphasis on student agency in contemporary education practices and research (Stenalt & Lassesen, 2022). Mameli and Passini (2019) attributed this to a substantial tension between students' free and authentic participation and teachers' dominant roles in kindergarten through grade 12 and higher education classrooms. Through a conceptual review, Matusov et al. (2016) mapped four conceptual frameworks of approaching student agency in educational studies: instrumental agency (human mastery of skills, knowledge, and dispositions to achieve predefined goals), effortful agency (the ability to act despite internal and external resistance, emphasizing self-regulation and commitment to sustain energy and motivation toward goals), dynamically emergent agency (agency arises from complex, interconnected processes involving human and nonhuman elements), and authorial agency (the ability to define one's own path by creating and shaping culture rather than conforming to existing norms).

In the context of AI-assisted learning, we adopt the effortful agency perspective because it foregrounds the learner's capacity to sustain intentional, self-aware engagement with tasks despite the convenience, automation, and potential cognitive offloading offered by GenAI tools. This perspective highlights the importance of learners actively resisting the temptation to delegate complex thinking to GenAI tools, thereby maintaining motivation, critical evaluation, and ownership of their learning processes in environments in which human effort can be easily displaced (Franzese, 2013). It aligns closely with Reeve's (2013) conceptualization of agentic engagement, which situates student agency within a broader engagement framework. Reeve positions agentic engagement as the fourth dimension of student engagement, alongside behavioral, cognitive, and emotional engagement, offering a concrete anchor for understanding how learners proactively contribute to and shape their own learning experiences (Reeve & Shin, 2020; Reeve & Tseng, 2011).

Reeve (2013) further delineated two complementary natures of agentic engagement: proactive and transactional. The proactive nature refers to learners' tendency to act in anticipation of activities delivered by others (e.g., teachers), actively shaping the unfolding of their educational experiences rather than simply reacting to predetermined tasks (Reeve & Tseng, 2011). In AI-assisted learning, this means students do not passively accept AI-generated outputs but instead initiate purposeful interactions—such as crafting targeted prompts, seeking clarifications, or requesting alternative perspectives—that steer the AI toward meeting their learning goals. The transactional nature frames agentic engagement as an ongoing series of dialectical exchanges between learners and others, in which each party's actions influence the other over time (Reeve, 2013). Traditionally examined in human–teacher interactions, this reciprocity can be extended to human–AI contexts, in which students' inputs guide the AI's responses, and the AI's outputs, in turn, shape subsequent student decisions (Kim, 2022; Reeve & Shin, 2020). Such reciprocal causation captures the dynamic interplay in which both human and AI contributions co-evolve, influencing the trajectory and quality of the learning process.

Guided by the agentic engagement theory, we propose the following research question for a grounded theory study:

How do higher education students manifest agency in their interactions with GenAI tools within AI-assisted learning environments?

## Methods

### Research Contexts and Participants

This study was conducted at a research-oriented university in Australia. When the data collection took place in July 2023, the university had issued a GenAI policy, including general guidance and best practices, to guide the ethical use of GenAI tools for faculty and students. The university also adopted Turnitin for detecting AI usage and plagiarism in student assignments. To recruit research subjects, recruitment emails were sent to potential participants through institutional channels, containing key information about our study, including research aims, participation procedures, and research ethics. From the pool of candidates, we employed a purposive sampling method to select the subjects with three criteria: (a) The students should have used certain GenAI tools, such as ChatGPT, for at least 4 months on a daily basis; (b) diversity in majors and years of study; and (c) gender balance. A total of 26 students were enrolled as research subjects, with 13 from STEM majors and the rest from humanities and social sciences, including arts, business, education, linguistics, and sociology. All participants were compensated with AUD60 for their participation in the study. Informed consent was obtained from all participants prior to data collection.

## Data Collection

Two types of data were collected from the research participants: conversation records with GenAI tools and cognitive interviews. Participants were first asked to submit at least five chat history records generated using specific GenAI tools that they had created during the learning activities. Each record had to include a minimum of five question-and-answer exchanges between the students and the GenAI tool. Records were collected via shareable links or exported PDF files to preserve the original format of the interactions. These conversation records provided authentic and situated artifacts for observing and analyzing students' behaviors when engaging with GenAI tools.

Prior to each interview, two researchers independently reviewed the participant's conversation records to familiarize themselves with the interaction style and identify key moments, patterns, and themes. Particular attention was paid to instances that might reflect agency, such as prompt modification, iterative questioning, or efforts to guide and correct the AI's responses. These flagged moments were then used to inform the cognitive interviews, enabling focused exploration of students' decision-making processes, intentions, and reflections during their interactions with GenAI tools.

Following the review of conversation records, cognitive interviews were conducted to examine each participant's interactions with GenAI tools in greater depth. A cognitive interview is a qualitative research technique used to investigate how individuals understand, interpret, and respond to tasks or situations by eliciting their underlying thought processes and reasoning (Willis, 2004). In this study, participants were shown selected excerpts from their GenAI conversation records as a stimulus and invited to reflect on their intentions, decisions, and strategies in those moments—not only what they did but also why they did it—in a think-aloud manner (Leighton, 2017). Example questions included the following: "Why did you start with this question or prompt?" "Were you satisfied with the response? Why or why not?" and "How would you describe your overall interaction process with ChatGPT?" This approach enabled the researchers to capture the cognitive and decision-making processes underpinning participants' actions, offering deeper insight into how they exercised agency in real time. Each interview lasted 1½ to 2 hours and was video-recorded and transcribed verbatim for analysis.

## Data Analysis

A grounded theory approach was adopted to analyze the collected data and develop a theoretical framework of student agency in AI-assisted learning (Glaser &

Strauss, 2017). The analysis began with open coding, in which the researchers closely examined the conversation records and interview transcripts line by line (Glaser, 2016). The data were broken down into discrete parts, and initial codes were created and assigned to segments of text that captured the specific actions, thoughts, and decisions made by the participants. This process identified and labeled emerging patterns and categories about students' strategies and actions behind their agentic engagement. Next, during axial coding, these initial codes were grouped into broader categories based on their similarities, differences, and the relationships between these categories. It helped to refine and connect the initial codes into more coherent subdimensions, revealing the underlying structure of student agency. For instance, behaviors related to modifying prompts in response to AI outputs were categorized under "adaptive refinement."

In the selective coding phase, the researchers focused on identifying core categories that best represented the central aspects of student agency. The analysis concentrated on integrating and refining the identified categories to develop a cohesive theoretical framework. These core categories were synthesized into four key aspects of agentic engagement, each further broken down into specific subdimensions. The final step involved theoretical integration, in which these aspects were woven into a cohesive framework that reflects the dynamic and iterative nature of student agency in AI-assisted learning. The framework was then validated by revisiting the data to ensure that all relevant instances of student agency were captured and accurately represented. This iterative process also involved peer debriefing and member checking, in which initial findings were discussed with other researchers and selected participants, to verify the accuracy and relevance of the identified categories and their relationships (McKim, 2023).

## Results

The data analysis revealed four aspects of student agency in AI-assisted learning activities: control and (re)direct, mindful adoption, external help-seeking, and reflective learning. These four aspects of agentic engagement did not take place linearly but, rather, iteratively and dynamically throughout the learning process. Students moved fluidly between these aspects, adapting their strategies in response to the challenges and opportunities they encountered while working with AI tools. In this section, these four aspects are elaborated with detailed evidence.

### Initiating and (Re)Directing

Students demonstrated agency by actively guiding and directing AI tools to ensure that the generated outputs aligned with their specific learning needs. This aspect was primarily reflected in their prompts and operations—how students composed and adjusted their instructions and other settings to steer the GenAI tool toward producing outputs that addressed the specific requirements of their tasks and met their learning objectives. It emphasizes deliberate control over the GenAI's functions to optimize its contributions to academic tasks and fulfill learning needs. This agency includes two subcategories: intentional initiative and adaptive refinement.

### *Intentional Initiative*

Intentional initiative refers to students' proactive and deliberate actions in guiding GenAI tools to generate content or perform tasks in alignment with their learning goals. Central to this was a proactive goal-setting, in which students established clear and explicit objectives for their interactions with AI. These well-defined goals served as the foundation upon which students composed and tailored their prompts, ensuring that the AI- generated outputs were specifically designed to meet their academic needs. Students' intentional initiative could be best demonstrated by articulating their requests, tasks at hand, or background information. For example, one student named John (a pseudonym), who was working on a product design project for his engineering design course, provided a list of functional requirements for a wallet design, as shown in Figure 1. Rather than a general requirement for wallet design, John explained that a detailed list of functional requirements was essential to obtain relevant and useful suggestions from ChatGPT. In this way, he exercised control over ChatGPT's task performance by taking deliberate steps to shape its outputs in a way that supported his academic and practical goals.

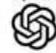

**Figure 1** Screenshot of Conversation Records Between John and ChatGPT

This intentional initiative was not limited to setting goals for AI but also reflected in students' deliberate actions in composing prompts and configuring settings of the AI systems to fulfill their learning goals. They actively leveraged their understanding of the subject matter and GenAI's potential to craft prompts that would yield the most beneficial results for their academic tasks. For example, many students employed techniques such as the "act as" prompt, where they assigned specific roles to GenAI to elicit more contextually appropriate and relevant responses. In addition, they actively manipulated the settings of their interactions with AI. For example, one student deliberately started new conversation threads to reset the context of interactions so that ChatGPT's responses were not influenced by previous exchanges. This careful management of AI's memory and contextual awareness shows students' active role in guiding AI to meet their specific educational needs, further illustrating the depth of their agency in AI-assisted learning.

*Adaptive Refinement*

This subcategory focuses on the iterative, adaptive process of adjusting and modifying prompts in response to AI-generated outputs, allowing students to continuously refine AI's contributions to better align with their evolving academic needs. This adaptive refinement involves students actively guiding the AI to generate or regenerate content that becomes increasingly specific and relevant as the interaction progresses. One common strategy observed was the approach of narrowing down topics from general to specific. Students began by providing GenAI tools with a broad topic and then progressively focused the conversation on more particular aspects of their primary concern. For example, one student started with the prompt, "What is the relationship between art and AI?" and then refined the inquiry by asking, "How do I combine AI and art design in student projects?" This sequential narrowing allowed the student to explore different facets of the subject matter, ultimately obtaining a more tailored and relevant output.

Moreover, when AI's responses deviated from the students' intended goals, they took proactive steps to correct the course of the conversation. This process often involved pointing out mistakes in the generated outputs and rephrasing or refining prompts to steer the AI back on track. In some cases, students described this process as a negotiation, in which they would iteratively adjust their instructions until the AI provided a satisfactory response. One student likened this to creating a "correct" environment for ChatGPT, in which repeated corrections and clarifications led to progressively more accurate outputs. Such iterative and adaptive interactions required effort, patience, and persistence, where students actively engaged in a trial-and-error process to achieve their desired outcomes. In essence, adaptive refinement exemplifies the dynamic and responsive nature of student agency in AI-assisted learning. Through persistence and strategic adjustments, students sought to harness the capabilities of GenAI tools more effectively and better meet their learning needs.

## Mindful Adoption

Whereas the initiating and (re)directing aspect focuses on students instructing and setting AI to generate content, the mindful adoption aspect shifts the focus to how they engage with and utilize the generated content. Mindful adoption involves students' critical and selective engagement with generated content, ensuring that the content they adopt to integrate into their work is of high quality, relevant, and purposeful. This aspect highlights the importance of thoughtful decision-making in utilizing AI outputs. This aspect includes two subcategories.

### *Critical Evaluation*

Upon receiving AI-generated outputs, agentic students tended to rigorously assess the outputs to determine their accuracy, relevance, and quality before incorporating them into their work. This process involved a careful examination of how well the AI's responses aligned with the original intent of their queries and whether the outputs genuinely contributed to their academic objectives. For example, in the previously discussed case of John, he carefully observed and evaluated the AI-generated output in Figure 1 as follows:

> *First, it's definitely not at the level of a functional requirement. For example, my first need was that I wanted a wallet that feels smooth, and it just responded with "smooth texture." That's basically just paraphrasing my user need. Similarly, when I mentioned "easy to remove," it translated that into "easy removal" as a functional requirement.*

In the evaluation, John recognized that ChatGPT's responses, although seemingly aligned with his input, lacked the depth and specificity required for true functional requirements. This critical reflection allowed John to discern that ChatGPT's output, although technically correct, did not provide the level of detail necessary to advance his design project meaningfully. This evaluation process illustrates a sophisticated level of critical engagement with AI-generated content, in which he did not simply accept the outputs at face value but, rather, analyzed their utility and relevance to his specific academic needs.

Other students had their own criteria in evaluating AI outputs. For example, another student emphasized the importance of relevance in the evaluation process. She noticed that although Claude's answers seemed correct on the surface, they often lacked depth or substantial usefulness. Such observations prompted her to consider relevance as the top priority, followed by accuracy and detail. This reflective process demonstrates how students evaluated the utility of AI-generated content, ensuring that what they adopted was not only superficially correct but also meaningful and applicable to their specific academic context. Such careful scrutiny is essential for mindful adoption, in which students exercised agency by filtering AI outputs through a lens of academic rigor and practical applicability.

### *Contextual Integration*

After critically evaluating AI-generated content, the students often identified suitable content and integrated it into their specific learning tasks in a careful, cautious, and contextually appropriate manner. This process, which we term "contextual integration," involves weighing the value and relevance of elements in the AI- generated content and thoughtfully incorporating them into their work. Students exercised this contextual integration by balancing the strengths and limitations of AI's outputs; they also made necessary adjustments to the content and aligned it with the specific context of their tasks. Such thoughtful actions are exemplified in a student's analysis of AI responses in a policy analysis task:

> *It [the AI tool] gave me a very structured response. It started from a broad perspective and provided several angles for thinking. However, I still had to work out the specific details on my own [even though it did provide some of the details] I feel that the data and details it provided are very likely to be biased, but the structure itself was correct. For these structured elements, I tend to adopt them because AI is indeed good at this, but for sure, I need to modify the details within that structure.*

The student's approach highlights a nuanced understanding of how to leverage GenAI tools effectively— acknowledging their strengths in generating broad, structured frameworks while also recognizing the need for human intervention to ensure accuracy and contextual relevance. This kind of agentic engagement was also reflected in students' cautious discernment in deciding when not to use AI tools for certain tasks. For example, one student mentioned that she would never use ChatGPT to process tasks relevant to literature reviews and public data sets, due to the potential for made-up or inaccurate information in its generated content. By selectively using GenAI tools, students can capitalize on AI's capabilities while still maintaining control over the finer details of their academic work. This process is indicative of a thoughtful, responsible, and strategic approach to engaging with AI, in which students actively curate the content they choose to integrate, ensuring its quality and alignment with their learning goals.

**External Help-Seeking**

Even within the context of AI-assisted learning, student engagement often extended beyond the GenAI tools they were using to include external resources and sources of help, especially among active and agentic students. External help-seeking refers to students' proactive efforts to supplement, validate, or enhance their AI-assisted learning by utilizing additional educational resources and human expertise. This kind of agency became crucial when students recognized the limitations of AI, which cannot fully substitute the breadth and depth of human expertise or conventional resources. As a result, they sought out additional support, demonstrating their ability to actively navigate complex information landscapes and take full ownership of their learning outcomes. This aspect involves two key subcategories.

*Supplementary Information-Seeking*

Supplementary information-seeking is concerned with the process by which students actively seek out additional information and resources from external platforms—such as textbooks, academic databases, or specialized websites—to complement and validate the outputs provided by GenAI tools. This subcategory reflects students' recognition of the limitations inherent in AI-generated content and their commitment to ensuring the accuracy, reliability, and depth of their learning outcomes. For instance, many students described their practice of double-checking the content generated by GenAI tools when writing essays. Although they appreciated the quick responses provided by AI, they emphasized the importance of consulting more authoritative sources, such as authoritative websites and Google Scholar, to verify the accuracy and relevance of AI's outputs. They noted that relying solely on AI was insufficient for producing high-quality academic work.

This example illustrates how supplementary sourcing is not just about fact-checking but also about enhancing the quality of academic outputs through the integration of traditional learning methods. Moreover, it signifies a deeper sense of responsibility, as students take ownership of their learning by not merely accepting AI outputs at face value but, rather, actively engaging in a process of verification and enhancement.

### *Collaborative Inquiry*

Collaborative inquiry took place when students actively and purposefully sought support from peers, instructors, and other knowledgeable individuals to enhance their learning process. Recognizing the limitations of AI, these students understood the value of leveraging human expertise to gain deeper insights, clarify uncertainties, and explore alternative perspectives that AI tools could not provide. This form of help- seeking was characterized by intentional interaction and dialogue, where students shared unresolved problems and engaged in discussions that tapped into the collective wisdom of others. By doing so, they made full use of the available resources around them, acknowledging the unique strengths that interpersonal relationships and human interactions bring to the educational experience. For example, one student described feedback and guidance from her academic advisor as irreplaceable in the interview:

> *Sometimes, even after trying different prompts, the help I get from ChatGPT is just average. It's not my fault, and it's not GPT's fault either—the core issue is that GPT can't handle every task, because no person or tool can do everything. If I can't sort it out myself, my first priority is to ask my professors for help. Often, just a few words from them can solve problems I've been struggling with for days.*

This interview demonstrates the student's ability to evaluate when to rely on AI and when to seek human support, as well as a proactive approach to managing his learning experiences. By purposefully combining the strengths of AI and human expertise, students embarked on a collective inquiry journey to achieve their personal objectives and enhance their learning experiences. This strategic approach to help-seeking also demonstrates their persistence, commitment, and responsibility with their own learning, reflecting a high level of agency. In this regard, collective inquiry is not just about seeking help but also about engaging in a purposeful and strategic process to enhance the overall quality of learning. It reflects a holistic approach to education in which students actively manage emerging challenges and take control of their education journey through informed decision-making.

## Reflective Learning

In addition to specific tasks, students' agentic learning process also involved ongoing reflection and self- scrutiny during and after their interactions with AI tools. In this reflective learning process, students critically analyzed the effectiveness of their AI use and their own actions, through which they identified opportunities for improvement in future interactions. In this regard, reflective learning enabled students to actively and intentionally refine and improve their AI-assisted learning experiences in a self-directed manner. This process not only reinforced student autonomy but also empowered them to take a more adaptive and resilient approach to their personal development. This aspect of student agency can be approached from two subcategories.

### *Process Reflection*

Process reflection is concerned with a regular and critical evaluation of students' manipulation of GenAI tools to identify strengths and areas for improvement in their actions. This reflection focused on the immediate, tactical aspects of using GenAI tools, such as the prompts used, the technical settings configured, and the overall approach to using AI in academic tasks. The goal of process reflection was to derive actionable insights and best practices that could enhance the effectiveness and quality of future interactions with GenAI tools. For example, many students recalled how they changed and adjusted their prompts through process reflection, and eventually they recognized prompt types or questioning strategies that could yield the most relevant and high- quality responses from AI.

In addition to simply refining their prompt strategies, process reflection also led students to a deeper understanding of the capabilities and limitations of AI, prompting them to adjust their expectations accordingly. One student commented, "I think realizing what AI cannot do is more important." In this regard, process reflection served as a mechanism for students to bridge the gap between their expectations and the actual outcomes they achieved via AI, by continuously improving and refining their technical proficiency. From this perspective, student agency is not merely about encouraging active and proactive behaviors; it also involves cultivating an informed and rational mindset that allows students to maintain control over their learning processes. Through process reflection, students develop the capacity to not only act but also act wisely, making decisions that are grounded in a deep understanding of both the potential and the limitations of AI.

### *Metacognitive Reflection*

Metacognition generally refers to thinking about one's thinking, and metacognitive reflection in AI-assisted learning is concerned with critically examining how students' cognitive processes interact with GenAI tools and the implications of these interactions for their learning. This form of reflection moved beyond the immediate, task-oriented evaluation of AI use in process reflection, delving into a more profound assessment of how AI influenced their understanding, decision-making, and approach to learning. By engaging in metacognitive reflection, students developed a heightened awareness of their cognitive strengths and limitations, recognized the biases and constraints inherent in AI-generated outputs, and made informed decisions about when and how to effectively incorporate AI into their academic endeavors. For example, one student reflected on the impact of AI use on her creative thinking:

> *I realized that I was relying too much on ChatGPT for generating ideas, and it was starting to limit my creativity. I had to step back and think about when to use it and when to push myself to come up with ideas independently.*

This reflection shows the student's awareness of the subtle ways in which AI could potentially limit her cognitive development, particularly in areas that require originality and critical thinking. By consciously deciding to limit her use of AI in specific contexts, the student exercised metacognitive control: She sought to preserve her cognitive autonomy while remaining an active agent in her learning process. Another student articulated a similar metacognitive realization, in which she questioned whether her reliance on AI had overridden her own judgment. Metacognitive reflection, therefore, enabled students to make conscious choices about how and when to utilize AI in ways that enhance, rather than detract from, their cognitive development. This form of reflection embodies self-regulation, critical thinking, and strategic decision-making, all of which empower students to shape and direct their own learning experiences.

## Discussion and Conclusion

Through a grounded theory approach, this study examined how students exercised agency in AI-assisted learning and identified four types of agentic engagement: initiating and (re)directing, mindful adoption, external help-seeking, and reflective learning. Figure 2 presents a visual representation of the theoretical framework of agentic engagement in AI-assisted learning. As shown in Figure 2, these four aspects show how students mindfully engaged with AI, made strategic decisions about their AI use, and integrated external resources to enhance their learning experiences. They further uncover how students exercised control, adapted to new technologies, and reflected on their learning processes, ultimately empowering them to navigate the complexities of AI-assisted learning with greater autonomy and effectiveness. Together, these four aspects contribute to a theoretical framework that captures the proactive and transactional nature of students' agentic engagement in AI-assisted learning. This framework highlights students' proactive, deliberate, and intentional efforts in managing their educational tools, resources, and experiences, through which they take the initiative, responsibility, and ownership of their own learning.

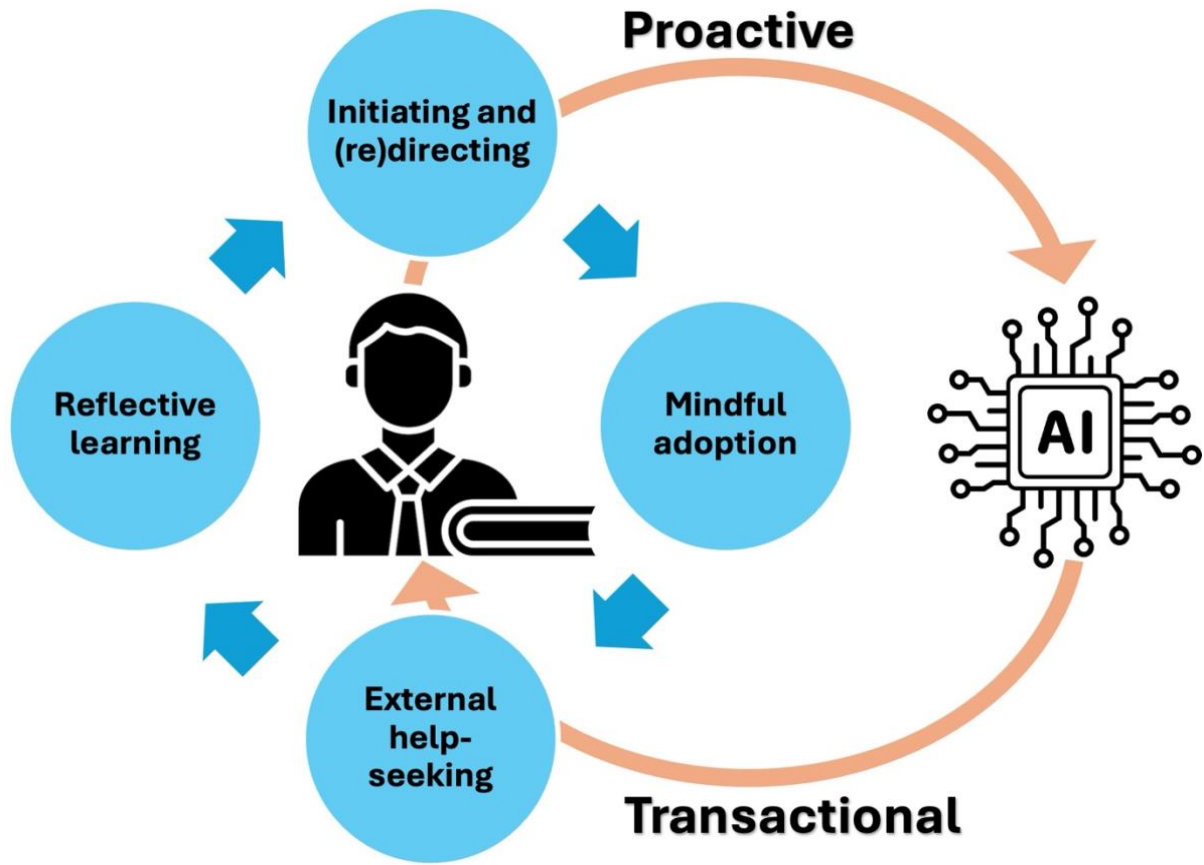

**Figure 2** Visual Representation of the Theoretical Framework of Agentic Engagement in AI-Assisted Learning

This study makes significant theoretical contributions by addressing a research gap in the literature of student agency, particularly within the context of AI-assisted learning. Historically, the concept of agency has been characterized by ambiguity, often approached through proxy measures rather than direct observation (Archer, 2000; Matusov et al., 2016). By identifying and articulating four distinct dimensions, this research offers a clear and operationalizable definition of student agency tailored to the emerging context of AI integration in education. This detailed framework moves beyond abstract conceptualizations, enabling a more precise observation and analysis of agentic behaviors in AI-integrated educational environments. Moreover, this framework can serve as a foundation for developing new, context-specific measurement scales, enabling more direct and accurate assessments of student agency in AI-assisted settings (Mameli & Passini, 2019).

Practically, this study offers valuable insights for educators, instructional designers, and policymakers seeking to harness GenAI technologies to enhance learning experiences while promoting student agency. The theoretical framework can serve as a guide for educators to design and implement instructional strategies that encourage students to actively, carefully, and responsibly integrate GenAI tools into their educational experiences (Roe & Perkins, 2024). By integrating these aspects into teaching methodologies, educators can foster more engaging, autonomous, and effective learning environments that leverage GenAI's potential without diminishing the learner's active role. For instructional designers, the findings inform the development of AI tools and platforms that are responsive to and supportive of student agency (Darvishi et al., 2024).

Designing AI systems that allow for customization, prompt refinement, critical engagement, and reflection can empower students to utilize these technologies more effectively and responsibly. Policymakers can also draw on this research to formulate educational policies and frameworks that balance technological advancement with the cultivation of essential learner skills and dispositions (Dai et al., 2024). Collectively, these practical contributions support a strategic and balanced approach to incorporating AI assistance in education, one that advances academic achievement while nurturing empowered and self-directed learners.

This study has several limitations, highlighting potential directions for future research. First, the findings are context-specific, derived from a particular educational setting, which may limit the generalizability of the results to other contexts or technologies. The findings might also be influenced by cultural and institutional factors (Dai, 2025; Hernandez & Iyengar, 2001). Although the proposed framework offers valuable insights into AI-assisted learning, its relevance across diverse educational environments and sociocultural contexts requires further testing, refinement, and adaptation. Second, the study's qualitative focus, although yielding rich, in- depth accounts of student experiences, is based on a relatively small sample size and may not reflect the full spectrum of student engagement with AI. Future research should adopt interdisciplinary, mixed-methods approaches that combine qualitative depth with quantitative breadth, enabling broader comparisons across populations, contexts, and AI applications (Green et al., 2017). Third, the rapid evolution of AI technologies poses a limitation to the study's long-term relevance, as future advancements in AI tools could significantly alter how students exercise agency in learning. Future studies are needed to enrich our understanding of student agency and the interplay between students and AI tools in the evolving technological landscape.

## Acknowledgment

A preliminary version of this study was presented at the 69th Annual Meeting of the Comparative and International Education Society (CIES 2025) in Chicago.